\documentclass[aps,prl,showpacs,twocolumn,superscriptaddress,amsmath,amssymb,floatfix]{revtex4}
\usepackage{tabularx}
\usepackage{bm}
\usepackage{epsfig,subfigure}
\usepackage{multirow}
\usepackage{graphicx}
\usepackage{color}
\usepackage{amsfonts}
\usepackage{exscale}
\usepackage{amsbsy}
\usepackage[colorlinks,urlcolor=blue,linkcolor=blue,anchorcolor=blue,citecolor=blue]{hyperref}

\begin{document}
\title{Distinct Spin Liquids and Their Transitions in Spin-$1/2$ XXZ Kagome Antiferromagnets}
\author{Yin-Chen He}
\altaffiliation{Current address: Max Planck Institut f\"{u}r Physik Komplexer Systeme, N\"{o}thnitzer Str. 38, 01187 Dresden, Germany}
\affiliation{Department of Physics, State Key Laboratory of Surface Physics and Laboratory of Advanced Materials, Fudan University, Shanghai 200433, China}
\author{Yan Chen}
\affiliation{Department of Physics, State Key Laboratory of Surface Physics and Laboratory of Advanced Materials, Fudan University, Shanghai 200433, China}
\affiliation{Collaborative Innovation Center of Advanced Microstructures, Nanjing 210093, China}
\begin{abstract}
By using the density matrix renormalization group approach, we study  spin-liquid phases of spin-$1/2$ XXZ kagome antiferromagnets. We find that the emergence of the spin-liquid phase is independent of the anisotropy of the XXZ interaction. In particular, the two extreme limits---the Ising (a strong $S^z$ interaction) and the XY (zero $S^z$-interaction)---host the same spin-liquid phases as the isotropic Heisenberg model. Both a time-reversal-invariant spin liquid and a chiral spin liquid with spontaneous time-reversal symmetry breaking are obtained. We show that they evolve  continuously into each other by tuning the second- and the third-neighbor interactions. At last, we discuss  possible implications of our results for the nature of spin liquid in nearest neighbor XXZ kagome antiferromagnets, including the nearest-neighbor spin-$1/2$ kagome anti-ferromagnetic Heisenberg model.
\end{abstract}
\maketitle

Ever since Anderson proposed the concept of quantum spin liquid \cite{Anderson1973}, hunting for exotic spin liquids in frustrated magnets has become a long and challenging journey in condensed-matter physics \cite{Balents2010, Marston1991, Hastings2000, Read1991, Wang2006, Yan2011, Lu2011, Jiang2012, Depenbrock2012, Hermele2005,Isakov2011,Yang1993, Ran2007, Messio2012, He2014, ssgong13, Senthil2000, Moessner2001, Balents2002, Kalmeyer1987, Rokhsar1988, Iqbal2013,Wen1989, Hermele2009, Schroeter2007, nielsen2012, Bauer2013, Kitaev2003, Han2012, Xu2009, Punk2014}. The gapped spin liquid is also a prominent example of topological order \cite{Wen1990}, which hosts fractionalized quasiparticles obeying fractional statistics that in principle can be used to implement topological quantum computation \cite{Kitaev2003}. Among  many frustrated systems,  spin-$1/2$ kagome antiferromagnet and its realizations in materials such as Herbertsmithite is considered to be the most promising candidate for a spin liquid; the neutron scattering experiment in particular has provided smoking-gun evidence for the existence of spinons \cite{Han2012}.

The minimal model for kagome antiferromagnets is  spin-$1/2$ kagome antiferromagnets, i.e., nearest-neighbor kagome antiferromagnetic Heisenberg model (NNKAH), for which the ground state is still controversial after more than twenty years of research \cite{Marston1991,Hastings2000,Singh2007, Evenbly2010, Read1991,Iqbal2013, Wang2006, Yan2011, Lu2011, Jiang2012, Depenbrock2012, Yang1993, Ran2007, Messio2012}. Among various studies, the density matrix renormalization group (DMRG) \cite{DMRG} results provide strong evidence for a gapped spin-liquid ground state \cite{Yan2011} with topological order \cite{Depenbrock2012,Jiang2012}. Recently infinite DMRG \cite{McCulloch2008} has been proven powerful in studying topological order by calculating topological degenerate ground states and the corresponding modular matrix \cite{He2014,Bauer2013,Cincio2013, Zaletel2013, He2013}; however, it fails for kagome spin liquids \cite{He2013}, making their nature remain elusive. One the other hand, theoretical descriptions of kagome spin liquid, which are heavily based on spinon-parton construction, have provided many fruitful results \cite{Lu2011, Ran2007, Hastings2000, Wang2006, Punk2014}, but there remains a gap to unify the theories, numerics and experiments.

\begin{figure}[htbp]
\centering
\includegraphics[width=0.35\textwidth]{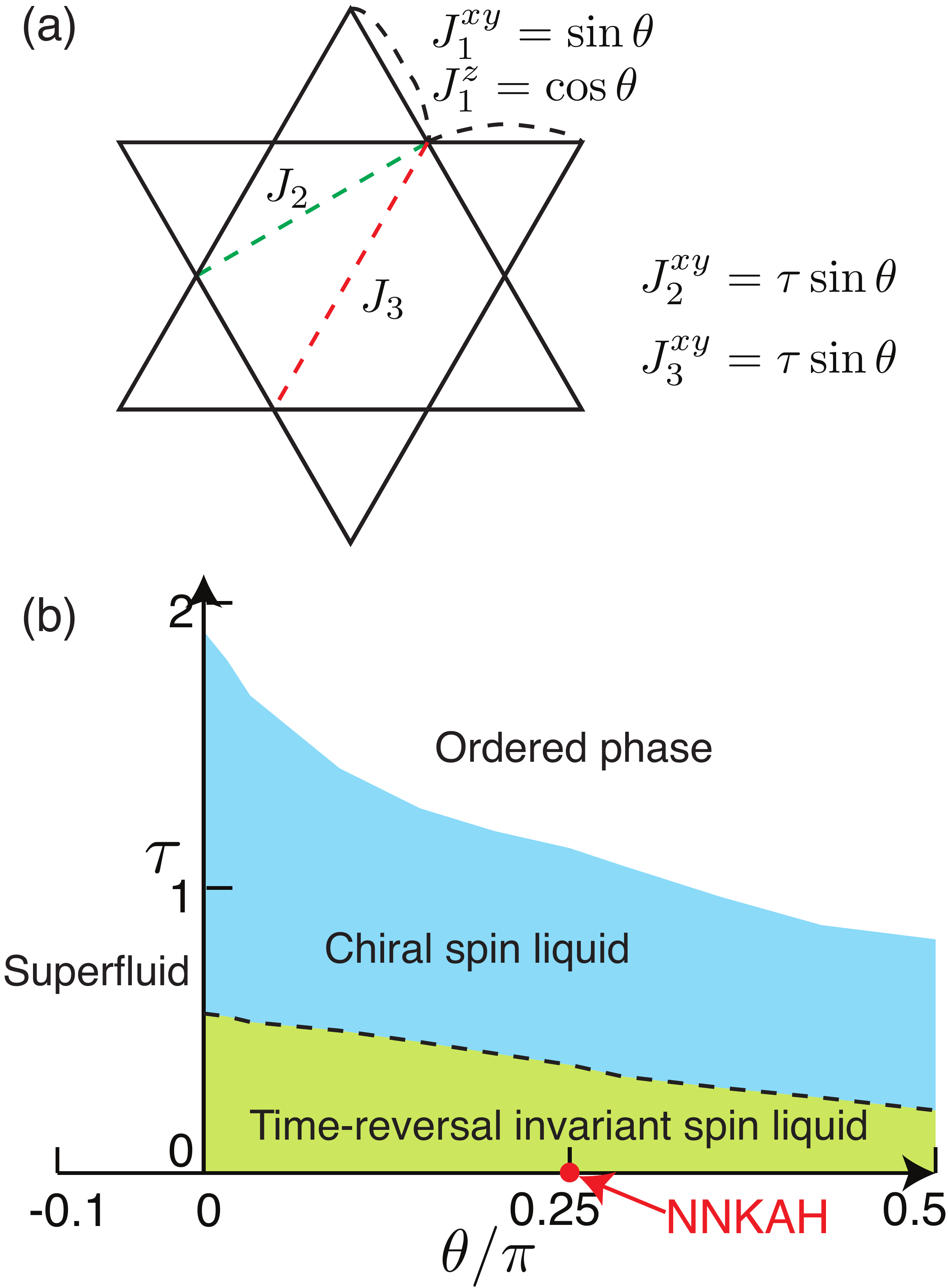}
\caption{\label{fig:phase_diagram} (Color online). (a) Kagome model with first- (XXZ), second-, and third- (XY) neighbor interactions. (b) Phase diagram of the kagome XXZ model.}
\end{figure}

In this Letter, we use the DMRG to study the spin-liquid phases in kagome antiferromagnets, where the $SU(2)$ Heisenberg-type spin-spin exchange interactions are extended into the XXZ type:
\begin{equation}
J\bm S_i \cdot \bm S_j \rightarrow J_z S_i^z S_j^z+J_{xy} (S_i^xS_j^x+S_i^y S_j^y).
\end{equation}
Besides the most studied $SU(2)$ limit ($J_z=J_{xy}$), we investigate two other limiting cases: Ising ($J_z\gg J_{xy}$) and XY ($J_z=0$) limits.  The underlying physics associated with these two limiting cases is very different from $SU(2)$ Heisenberg limit: the Ising limit can be mapped onto a quantum dimer model or compact $U(1)$ gauge field theory coupled with dynamical matter field \cite{Nikolic2005}; the XY limit  can be considered as a hard-core boson system, where a spin liquid state could be realized in an exotic way by fractionalizing \cite{Wang2014} (or fermionizing \cite{Vortexliquid}) vortices. Although these three limits are physically different, we find that the emergent spin-liquid phases (see Fig. \ref{fig:phase_diagram}) are almost the same (but $SU(2)$ Heisenberg point might have  richer ``symmetry enriched topological" phases \cite{SET, Wang2013} by the larger $SU(2)$ symmetry): (1) with only first-neighbor interactions (NNKAXXZ), the system hosts a time-reversal-invariant spin liquid that lies in the same phase as the spin liquid found in NNKAH ($SU(2)$ limit) \cite{Yan2011}; (2) with second and third neighbor interactions added \cite{He2014, ssgong13}, a chiral spin liquid (CSL) will emerge. Further more, we provide strong evidence for the continuous phase transition between those two spin liquid phases.

Our findings will  naturally bring new possibilities and questions about theoretically understanding of kagome spin liquid: which limit of kagome XXZ model is more fundamental for theoretical description of kagome spin liquid?  Should we continue to rely on the spinon-parton construction for describing kagome spin liquid or turn to approaches such as fractionalizing vortices \cite{Wang2014} in XY limit? On the other hand, the continuous phase transition between two spin liquids will help to narrow down the possibility of the candidate spin liquid phase realized in NNKAXXZ, including the most studied nearest neighbor kagome Heisenberg model.

\emph{Model Hamiltonian.}---Our model is a $J_1$-$J_2$-$J_3$ XXZ antiferromagnetic model defined on a kagome lattice, for which the Hamiltonian is:
\begin{align}
H=&J_1^z \sum_{\langle i j\rangle} S^z_i S_j^z+ J_1^{xy} \sum_{\langle i j\rangle} (S^x_i S_j^x+ S^y_i S_j^y)+ \nonumber\\ & J_{2}^{xy} \sum_{\langle\langle i j\rangle\rangle} (S^x_i S_j^x+ S^y_i S_j^y)+ J_{3}^{xy} \sum_{\langle\langle\langle i j\rangle\rangle\rangle} (S^x_i S_j^x+ S^y_i S_j^y),
\end{align}
where $\langle i j\rangle$ denotes first-neighbor, $\langle\langle i j\rangle\rangle$ second-neighbor, and $\langle\langle\langle i j\rangle\rangle\rangle$ third-neighbor interactions (see Fig. \ref{fig:phase_diagram}(a)), and we take $J_1^z=\cos \theta$, $J_1^{xy}=\sin \theta$, $J_2^{xy}=J_3^{xy}=\tau \sin \theta$. $\theta$ controls the anisotropy of the XXZ interaction: for $\theta\sim 0$, the system is in the Ising limit; for $\theta=\pi/4$, the system is in the $SU(2)$ limit; for $\theta=\pi/2$, the system corresponds to the XY limit. $\tau$ controls the relative magnitude of the second- and third-neighbor interactions.

We use the infinite DMRG algorithm \cite{McCulloch2008} to study the system wrapped on a cylinder with YC or XC geometry \cite{He2014}. We use a code with complex variables (CSL spontaneously breaks time-reversal symmetry), and keep 6000 states in the DMRG simulation (near the critical point, we have kept 10000 states). We obtain a generic phase diagram (Fig. \ref{fig:phase_diagram}(b)) based on the calculations on the YC8 ($L_y=4$ unit cells) cylinder. Except for the CSL phase, all other phases are time-reversal invariant: (1) with intermediate $\tau$ (second- and third-neighbor interactions), we have a CSL with spontaneous time-reversal symmetry breaking, as similarly studied in previous work \cite{He2014, ssgong13}; (2) for small $\tau$, we obtain a time-reversal-invariant spin liquid which lies in the same phase as the spin-liquid state found by Yan, \emph{et al}. \cite{Yan2011} (NNKAH); (3) with $\theta<0$ (ferromagnetic XY interaction), we get a superfluid phase, in agreement with quantum Monte Carlo results \cite{Isakov2006}; (4) for large $\tau$, an ordered phase is obtained \cite{footnote_order}.

\begin{figure}[htbp]
\centering
\includegraphics[width=0.45\textwidth]{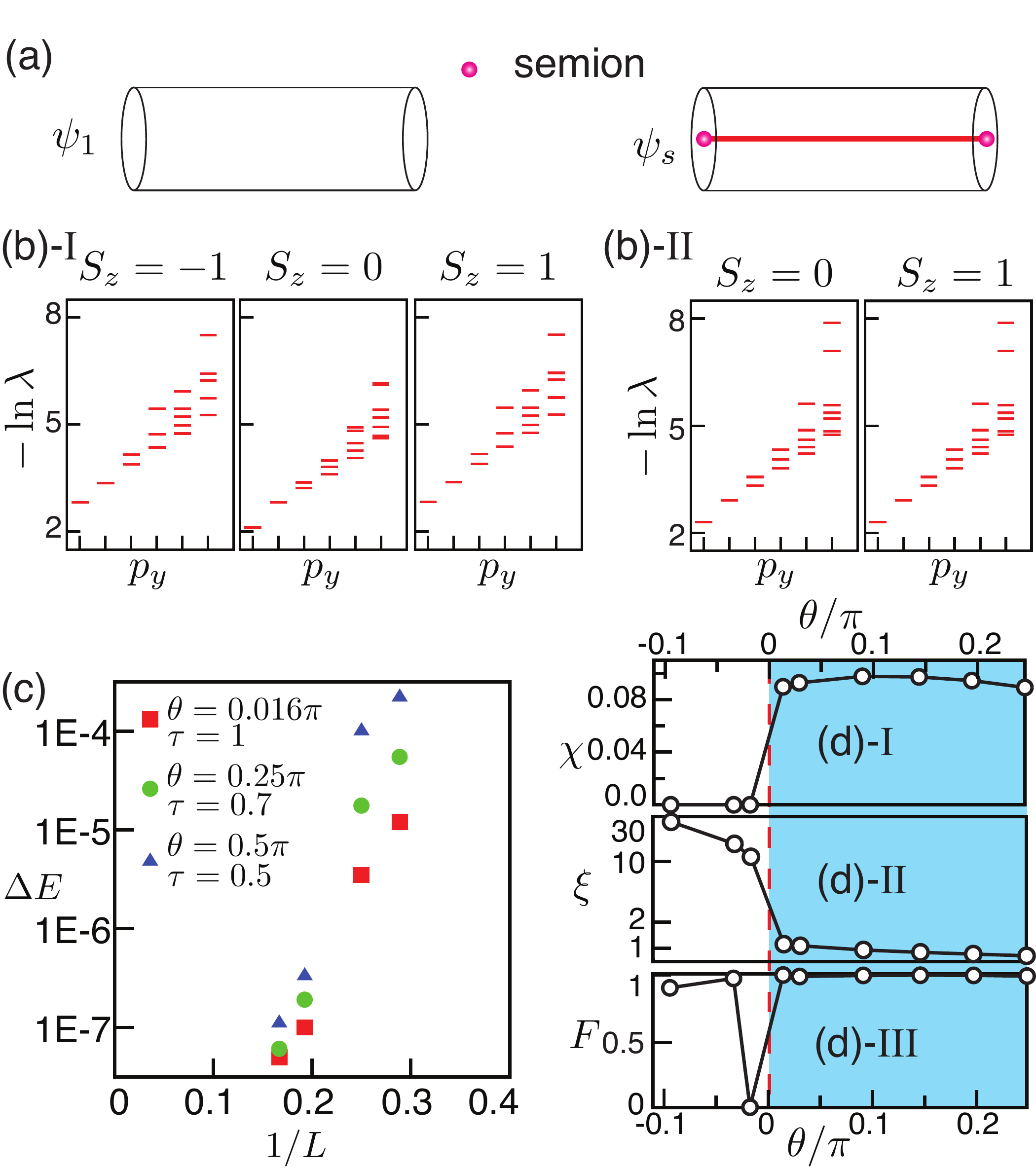}
\caption{\label{fig:CSL_XXZ} (Color online). (a) Two topological degenerate states of CSL. (b) Entanglement spectrum (sorted by quantum number $S_z$) of CSL on YC12 cylinder, here $\theta=0.016\pi$, $\tau=1$: (I) $\psi_1$ (II) $\psi_s$. The horizontal axis is the momentum along the $y$ direction, $p_y=0, 2\pi/6, \cdots, 5\times2\pi/6$ (up to a global shift). (c) Energy splitting between two topological degenerate ground states. (d) Transition from CSL ($\theta>0$) to superfluid ($\theta<0$), here we show results for the YC8 cylinder, $\tau=1$. I: Scalar chirality order parameter $\chi$; II: $\log$-plot of the correlation length $\xi$ (unit cells) extracted from the transfer matrix; III: The fidelity $F$ (defined in Eq. \ref{eq:fid}) between all pairs of neighboring points.}
\end{figure}

\emph{Chiral spin liquid.}---As shown in the phase diagram, the Hamiltonian hosts a CSL with spontaneous time-reversal symmetry breaking in certain parameter region, similar to previous findings \cite{He2014, ssgong13}. The CSL is a gapped spin liquid state that supports a fractionalized spinon-type quasiparticle---semion. The CSL has two topologically degenerate ground states, which can't be distinguished locally, but can be measured by global operator such as Wilson $W_y$ ($W_y=\pm 1$ for $\psi_1$ and $\psi_s$ up to a normalized factor), which winds a pair of semions around  the torus or infinite cylinder and annihilates them. There are two ways to get different topological sectors: (1) creating an semion line in $x$ direction, thus the semion winding around $y$ direction will perceive the semion line and gain a $-1$ phase (due to fractional statistics), then we have $W_y=-1$ for $\psi_s$. (2) inserting $2\pi$ flux in the system, the semion winding around $y$ direction of the torus will have an Aharonov--Bohm phase $\exp(i 2\pi c)=-1$ due to the fractional charge of semion. We have obtained these two ground states using the technique developed in Ref. \cite{He2013}, particularly from entanglement spectrum  (Fig. \ref{fig:CSL_XXZ}(b)) we can know $\psi_s$ has a semion line threaded in; also by inserting $2\pi$ flux, these two topological degenerate ground states will adiabatically evolve into each other. The degeneracy pattern ({1,1,2,3,5 $\cdots$ }) of the leading entanglement spectra in both ground states also agrees with the expectation of CSL state \cite{Li2008}. The energy splitting between the two  ground states decays exponentially fast (Fig. \ref{fig:CSL_XXZ}(c)), supporting the idea that the two states are degenerate in the thermodynamic limit.

Furthermore, using these two topological degenerate ground states obtained at various parameter points \cite{footnote2}, we calculate the modular matrix \cite{Wen1990,Zhang2012, Cincio2013}, which gives:
\begin{align}
\mathcal S&=\frac{1}{\sqrt 2}\left( \begin{matrix} 1 & 1 \\ 1 & -1 \end{matrix} \right)+ o(10^{-2})
\end{align}
and
\begin{align}
\mathcal U&=e^{-i (2\pi/24)} \left( \begin{matrix} 1 & 0 \\ 0 & i \end{matrix} \right)\times o(10^{-2})
\end{align}
From the modular matrix, we conclude that the system hosts a CSL. For example, from the $\mathcal S$ matrix, we know the fractional statistics obeyed by the semion: one semion encircling another semion will give rise to a non-trivial phase factor $-1$.

To show spontaneous time-reversal symmetry breaking, we measure the scalar chirality order parameter, $\chi=\langle \bm S_i \cdot (\bm S_j \times \bm S_k)\rangle$. As shown in Fig. \ref{fig:CSL_XXZ}(d)-I, the system has a large scalar chirality order parameter when $\theta>0$. For $\theta<0$, the system is in the superfluid phase, which is time-reversal invariant with a vanishing chirality order. From Fig. \ref{fig:CSL_XXZ}(d)-II, we find the correlation length (from the transfer matrix) \cite{McCulloch2008} is extremely large when $\theta<0$ (it should be infinite, but we get a finite value due to the truncation effect in DMRG), indicating a gapless superfluid state. In contrast, the correlation length for the CSL phase is very small signifying a gapped state. Finally, we calculate the ``one column'' fidelity, $F(\theta)$, for a cylinder (or torus) with length $L_x$ (large or infinite) and  width $L_y$  (finite and small), which is defined as:
\begin{equation}
|\langle \psi(\theta+\delta \theta)|  \psi(\theta)\rangle|= [F(\theta)]^{L_x} \label{eq:fid}
\end{equation}
This fidelity $F(\theta)$ can be easily calculated from the transfer matrix \cite{McCulloch2008}, and can serve as a good criterion for a nature of the phase transition. From the fidelity between $\theta>0$ and $\theta<0$ (Fig. \ref{fig:CSL_XXZ}(d)-III), we conclude the phase transition between the superfluid phase and the CSL is of the first order.

\emph{Time-reversal-invariant spin liquid.}---In the following, we consider the case $\tau=0$, where the Hamiltonian  has only nearest-neighbor interactions. If $\theta=\pi/4$, the interaction is the Heisenberg $SU(2)$ interaction (NNKAH), which according to DMRG simulations \cite{Yan2011,Depenbrock2012} is a time-reversal-invariant gapped spin liquid.

\begin{figure}[htbp]
\centering
\includegraphics[width=0.45\textwidth]{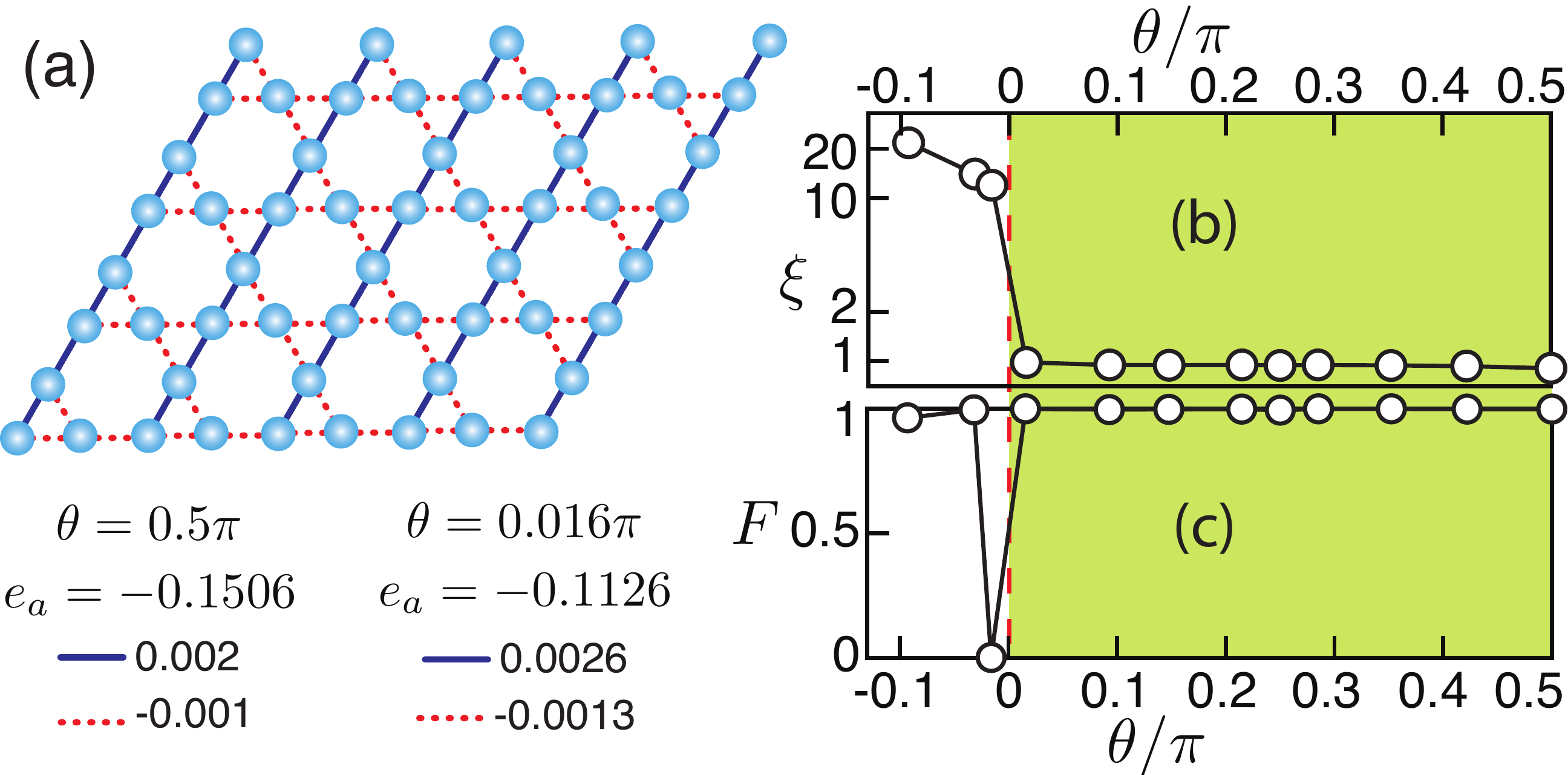}
\caption{\label{fig:XXZ_NN} (Color online). Results of NNKAXXZ ($\tau=0$) obtained on the YC8 cylinder. (a) Bond spin correlations along the XY direction, $\langle S_i^xS_j^x+S^y_iS_j^y\rangle -e_a$, ($e_a$ represents its average value). The transition from spin liquid ($\theta>0$) to superfluid ($\theta<0$). (b) log-plot of correlation length $\xi$ (measured in unit cells). (c) Fidelity $F$ (defined in Eq. \ref{eq:fid}) between all pairs of neighboring points.}
\end{figure}

The ground state does not break any lattice symmetry, in particular, the nearest-neighbor spin-spin correlations along XY direction ($S_i^xS_j^x+S^y_iS_j^y$) is very homogenous (Fig. \ref{fig:XXZ_NN}(a) presents two limiting cases---the Ising ($\theta=0.016\pi$) and the XY ($\theta=\pi/2$) limit). When $\theta>0$, the correlation length \cite{McCulloch2008} shown on Fig. \ref{fig:XXZ_NN}(b) is very small, and the fidelity (Eq. \ref{eq:fid}) between all pairs of neighboring points (Fig. \ref{fig:XXZ_NN}(c)) is approximately one. This result strongly suggest that there is no phase transition in the whole region of $\theta>0$, both the Ising and the XY limits being adiabatically connected to the spin liquid phase in the NNKAH \cite{Yan2011}. When $\theta<0$ (unfrustrated XY interactions), the system is in the gapless superfluid phase; the transition between superfluid phase and spin liquid phase in NNKAXXZ is of the first order.

\emph{Transitions between two spin liquids.}---Next, we study how the CSL evolves into the spin-liquid ground state in NNKAXXZ. Our results are based on the YC8 cylinder, the critical point may shift as the system gets larger or more states are retained in the DMRG simulation. However, the nature of the phase transition does not change. The spin liquid has topologically degenerate sectors; here we focus only on the vacuum sector (lowest energy state), which has a smaller finite-size effect \cite{He2014}.

\begin{figure}[htbp]
\centering\includegraphics[width=0.45\textwidth]{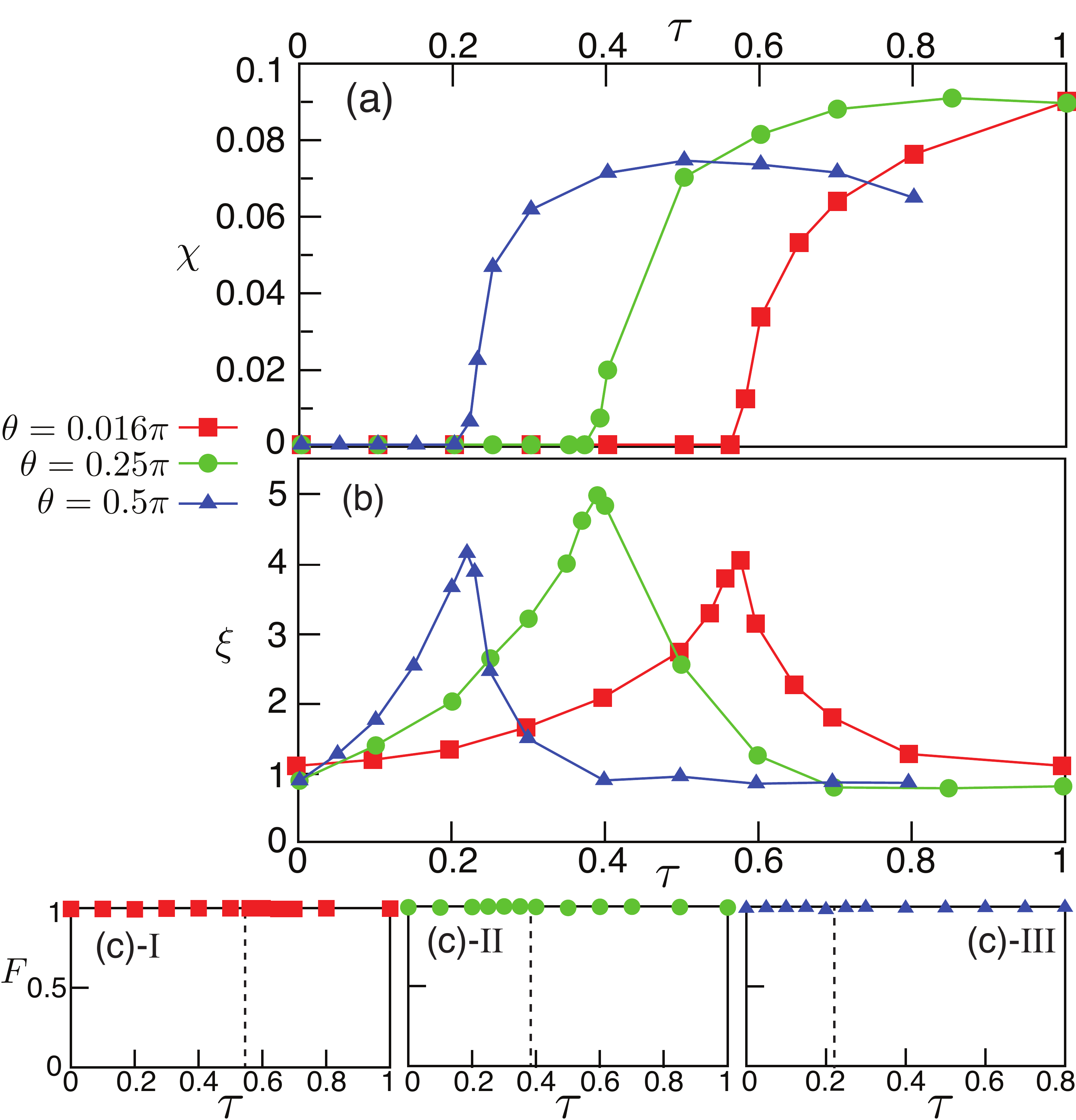}
\caption{\label{fig:CSL_NN}(Color online). Transition from CSL ($\tau\sim 1$) to spin liquid in NNKAXXZ ($\tau=0$); here we show results (YC8 cylinder) obtained in  the Ising ($\theta=0.016\pi$), the $SU(2)$ ($\theta=\pi/4$) and the XY ($\theta=\pi/2$) limits. (a) chirality order $\chi$, (b) correlation length $\xi$ (measured in unit cells), (c) fidelity $F(\tau)$ (Eq. \ref{eq:fid}) $F$ between pairs of neighboring points. The dashed line marks the critical point.}
\end{figure}

The CSL has spontaneous time-reversal symmetry breaking with finite scalar chirality order $\chi=\langle \bm S_i \cdot (\bm S_j \times \bm S_k)\rangle$ (on each triangle); hence, we use $\chi$ as the order parameter to distinguish the CSL and the time-reversal-invariant spin liquids. As shown in Fig. \ref{fig:CSL_NN}(a), for each $\theta$ there is a transition point after which the system enters into the time-reversal-invariant spin-liquid phase with vanishing chirality order. In the whole process, there is no spin rotational or lattice symmetry breaking, so we conclude there is no intermediate phase between the CSL and spin liquid in NNKAXXZ. The correlation length \cite{McCulloch2008} in Fig. \ref{fig:CSL_NN}(b) also reaches the peak at the critical point; it is consistent with a direct phase transition between CSL and spin liquid in NNKAXXZ. We remark that correlation lengths shown here have not been extrapolated with truncation error or the number of kept states. We find that near the critical point, the correlation lengths keep growing as more states are kept (see supplementary materials). They are supposed to be much larger, even infinite, if one retains more and more states.

To show unambiguously the continuous transition between the two spin liquids, we measure the fidelity $F(\tau)$ (Eq. \ref{eq:fid}) between pairs of nearest-neighbor points as $\tau$ varies progressively (Fig. \ref{fig:CSL_NN}(c)). The overlaps are very large ($\sim 0.99$) and hence clearly show that CSL continuously evolves into the spin-liquid state in NNKAXXZ without level crossing.

The continuous transition between CSL and spin liquid in NNKAXXZ actually narrows down the possibility of spin-liquid phase in such system. A recent work \cite{Barkeshli2013} has proposed an  interesting theory for the continuous phase transition between CSL and double-semion spin-liquid \cite{Freedman2004,Levin2003}. However, it is argued that \cite{Zaletel_dbs} double-semion phase cannot be realized in spin-1/2 kagome system (with $U(1)$ charge conservation), unless one enlarges the kagome unit cell or realizes the time-reversal symmetry in a twisted way \cite{Qi2014}. Neither of these phenomena have been observed in our numerical results, so the spin-liquid state obtained in our simulation appears not to belong to a double-semion phase.

A candidate for kagome spin liquid is a Z$_2$ spin liquid \cite{Jiang2012, Depenbrock2012}, particularly one may wonder whether our model in the Ising limit $\theta\sim 0$, $\tau=0$ ($J_1^z\gg J_1^{xy}$, $J_2^z=J_3^z=J_2^{xy}=J_3^{xy}=0$) can be connected to the Z$_2$ spin liquid proposed in Ref. \cite{Balents2002}, where one can have $J_1^z=J_2^z=J_3^z\gg J_1^{xy}$ \cite{Isakov2011}.  Numerically, we find that  these two phases won't be adiabatically connected as we simply change $J_2^z$, $J_3^z$: in the intermediate region ($J_1^z>J_2^z=J_3^z>0$), the system will enter ordered phases (valence bond solid). Also, we'd like to remark that the Z$_2$ spin liquid in Ref. \cite{Balents2002} is different from the usually discussed Z$_2$ spin liquid candidate for kagome \cite{Read1991, Lu2011, Wang2006} in the point of view of symmetry enriched topological phase \cite{Wang2013}: the spinon of Z$_2$ spin liquid in Ref. \cite{Balents2002} is Kramers singlet; the spinon in  Ref. \cite{Read1991, Lu2011, Wang2006} is Kramers doublet.  Another interesting question is whether one can have a critical theory for the transition from Z$_2$ spin liquid to CSL, which might be an exotic phase transition beyond Landau-Ginzburg-Wilson paradigm.

Besides the gapped topological spin liquid, U(1) Dirac spin liquid \cite{Ran2007, Iqbal2013} is also a promising candidate for the kagome spin liquid ground state. In particular,  by adding various mass terms, the U(1) Dirac spin liquid will continuously transform into CSL, valence bond solids, and magnetically ordered state \cite{Hastings2000, Hermele2005}, all of which have been found to be neighbors of  the kagome spin liquid. However, this U(1) Dirac spin liquid has a vanishing spin gap, which is inconsistent with the large spin gap reported in the DMRG's results \cite{Yan2011,Depenbrock2012}. A possible direction is to construct a new type of spin-liquid such as the fractionalized vortex liquid \cite{Wang2014} for the XY kagome  antiferromagnets, which might produce a critical spin liquid with a finite spin gap.

\emph{Conclusions.}---We numerically study the spin-liquid phases in spin-$1/2$ XXZ kagome antiferromagnets, and find both a time-reversal-invariant spin liquid and chiral spin liquid with spontaneous time-reversal symmetry breaking, whose emergence is independent of the anisotropy of XXZ interactions. Furthermore, we show the phase transition between the two spin liquids is continuous.  Finally, we discussed possible future directions in understanding the spin liquid phase of kagome antiferromagnets.

\emph{Acknowledgments.}--- For stimulating discussions, we thank M. Barkeshli, D. N. Sheng, Y. Qi, Y.-S. Wu, Z.-C. Gu, T. Xiang, M. Zaletel and R. B. Tao. This work was supported by the State Key Programs of China (Grant No. 2012CB921604), the National Natural Science Foundation of China (Grant Nos. 11274069, and 11474064).

\emph{Note added.}---Upon completion of this work, we became aware of a related work \cite{Lauchli}, which studies nearest-neighbor XXZ kagome model by exact diagonalization and reaches similar conclusions as our own.

\newpage

\onecolumngrid

\begin{center}
{\bf \large Supplementary Material of ``Distinct Spin Liquids and their Transitions in Spin-$1/2$ XXZ Kagome Antiferromagnets"}
\end{center}

\begin{figure}[h]
\centering
\includegraphics[width=0.88\textwidth]{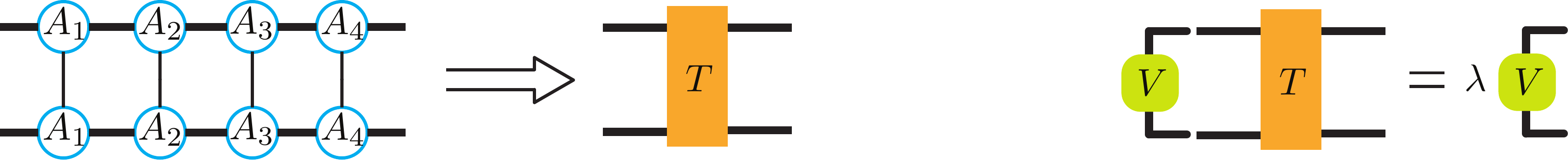} \caption{\label{fig:transfer}   Definition of transfer matrix and correlation length $\xi$.}
\end{figure}

To calculate the correlation length, we define the transfer matrix $T$ as in Fig. \ref{fig:transfer}. The correlation length is then defined by the first and second largest eigenvalue $\lambda_{1,2}$ of the transfer matrix $T$,
\begin{equation}
\xi_{\textrm{TM}}=-1/\ln \lambda_2
\end{equation}
This correlation length determines the largest correlation in the infinite cylinder \cite{McCulloch2008}. Therefore, instead of calculating various correlation functions, one can simply calculate this single quantity $\xi_{\textrm{TM}}$ to obtain the length scale of the largest possible correlations.

\begin{figure}[h]
\includegraphics[width=0.5\textwidth]{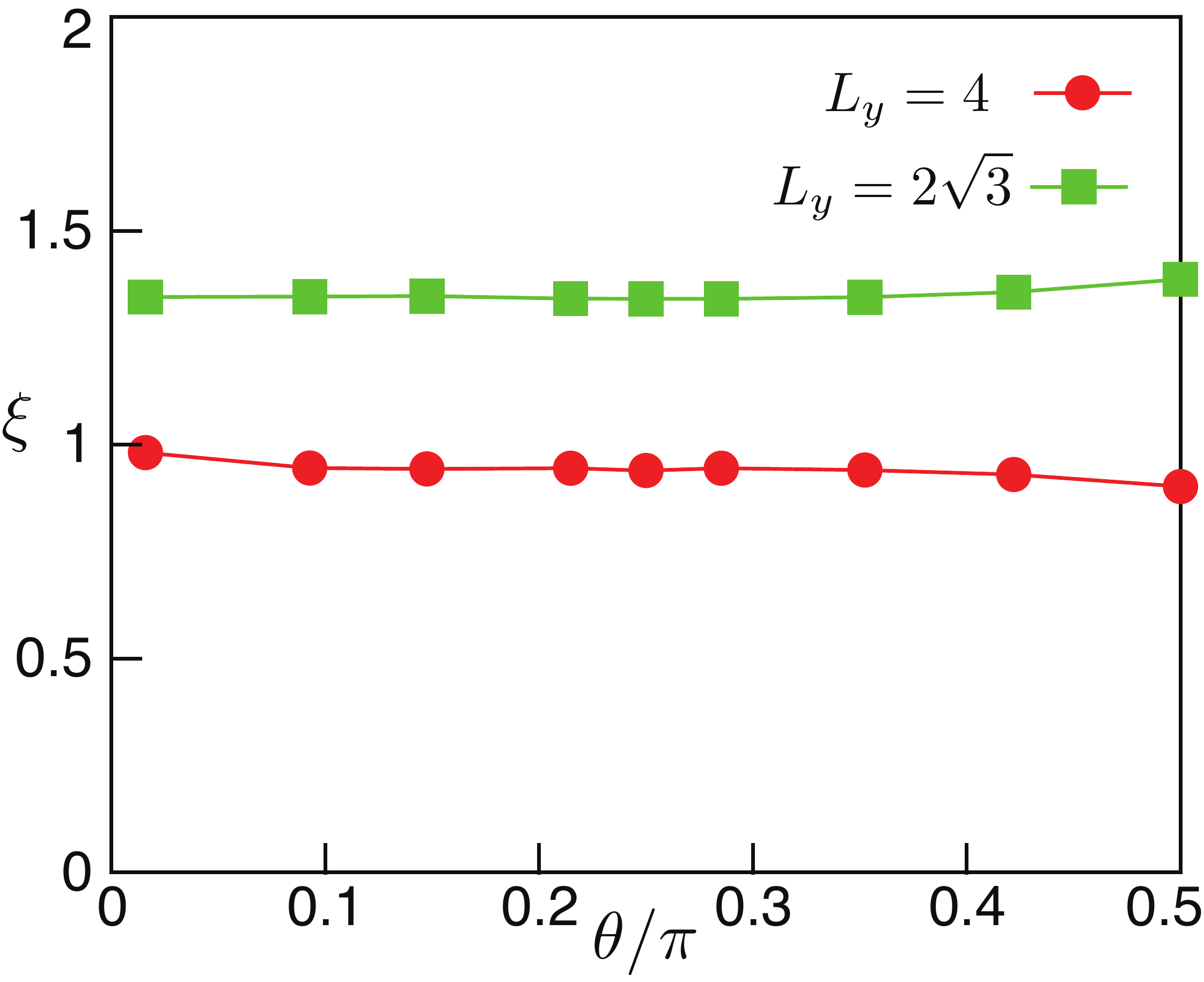}\caption{\label{fig:cor_vs_L}   Comparison of correlation length for two different system size ($\tau=0$), $L_y=2\sqrt 3$ unit cells (XC8) and $L_y=4$ unit cells (YC8). }
\end{figure}

Fig. \ref{fig:cor_vs_L} shows  how correlation lengths change as the system size varies, and the Hamiltonian we calculate only has first neighbor interactions ($\tau=0$). One can observe that for different system size, the correlation length behaves similarly as $\theta$ changes: they are all small for three different limits, and strongly suggest there is no phase transition between them.

\begin{figure}[h]	
\includegraphics[width=0.5\textwidth]{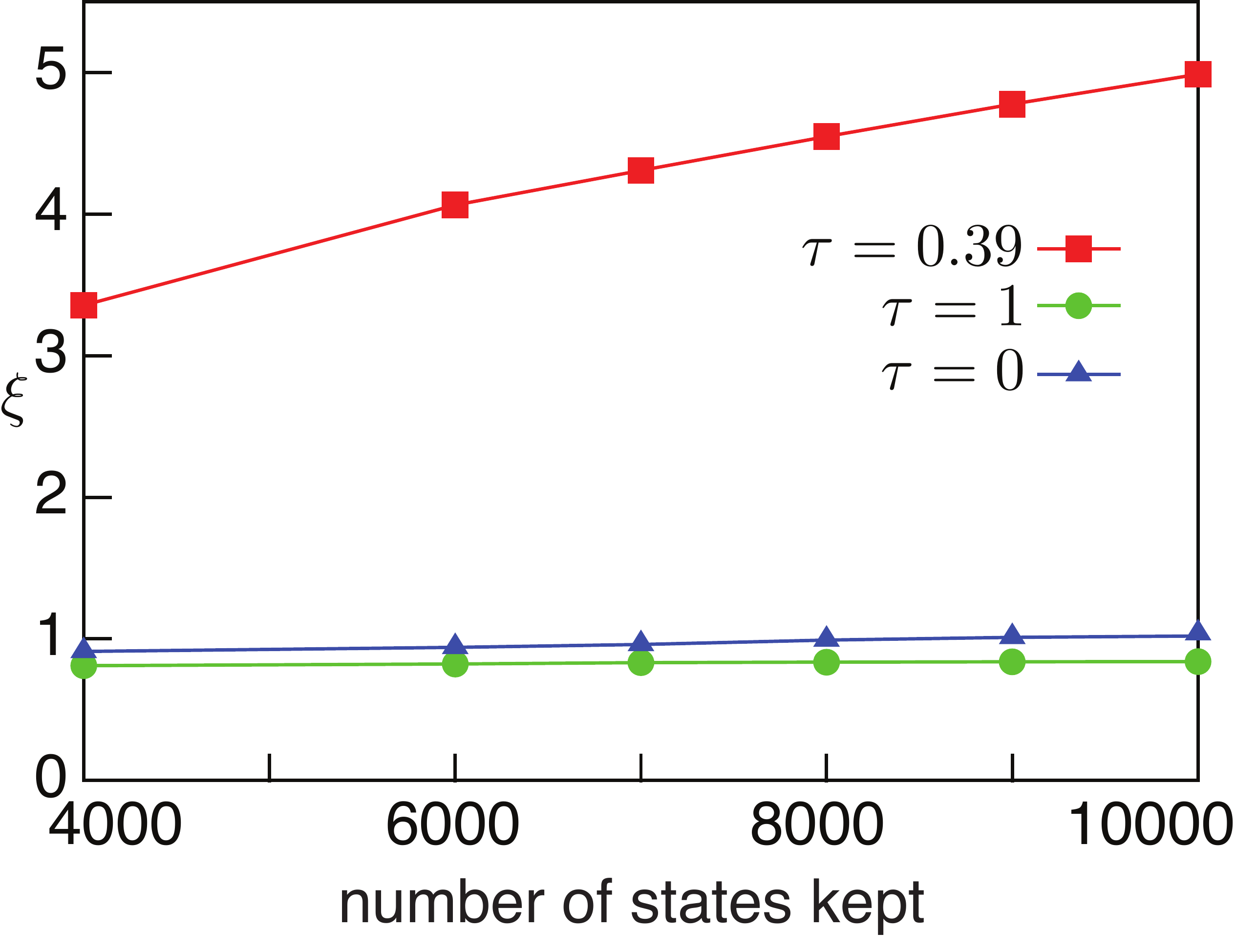}\caption{\label{fig:cor_vs_m} Correlation length versus the number of states kept. Here we show the Heisenberg limit $\theta=0.25\pi$.}
\end{figure}

Fig. \ref{fig:cor_vs_m} shows the correlation length as a function of the number of states retained. For the state that lies deeply in the spin liquid phase ($\tau=0,1$), the correlation fully converges. However, when the system is near the critical point ($\tau=0.39$), the correlation length does not converged for the largest number of states that we attempted.

\begin{thebibliography}{99}
\bibitem{footnote_order} The system seems to enter initially into a stripe phase, then a valence bond solid.
\bibitem{footnote2} The ground states we used are taken from all three limits, specifically Ising limit $\theta=0.016\pi$, $\tau=1$; $SU(2)$ limit $\theta=0.25\pi$, $\tau=0.7$; XY limit $\theta=0.5\pi$, $\tau=0.5$.
\end{thebibliography}
\end{document}